\documentclass[twoside]{article}
\usepackage{fleqn,espcrc2}

\usepackage{graphicx}
\usepackage{epsfig}

\newcommand{\AmS}{{\protect\the\textfont2
  A\kern-.1667em\lower.5ex\hbox{M}\kern-.125emS}}

\title{Microwave Electrodynamics of the Electron-Doped Cuprate Superconductors
Pr$_{2-x}$Ce$_{x}$CuO$_{4-y}$ and Nd$_{2-x}$Ce$_{x}$CuO$_{4-y}$}

\author{J. David Kokales, Patrick Fournier, Lucia V. Mercaldo, Vladimir V. Talanov, Richard L. Greene, and Steven M. Anlage\address{Center for Superconductivity Research, Department of Physics, \\ 
        University of Maryland, College Park, MD 20742-4111}}
       
\begin{document}

\begin{abstract}
The pairing state symmetry of the electron-doped cuprate superconductors is thought to be s-wave in nature, in contrast with their hole-doped counterparts which exhibit a d-wave symmetry.  We re-examine this issue based on recent improvements in our electron-doped materials and our measurement techniques.  We report microwave cavity perturbation measurements of the temperature dependence of the penetration depth of Pr$_{2-x}$Ce$_{x}$CuO$_{4-y}$ and Nd$_{2-x}$Ce$_{x}$CuO$_{4-y}$ crystals.  Our data strongly suggest that the pairing symmetry in these materials is not s-wave.
\vspace{1pc}
\end{abstract}
\maketitle

\section{Introduction}

Existing experimental evidence suggests that the pairing state symmetry in the electron-doped cuprate superconductors is s-wave in nature, in contrast to the d-wave symmetry observed in the hole-doped cuprates.  There is no compelling theoretical reason for this difference.  The strongest evidence for s-wave symmetry comes from measurements of the penetration depth \cite{Wu}.  However, it is possible that the paramagnetism of the Nd rare-earth ion in Nd$_{2-x}$Ce$_{x}$CuO$_{4-y}$(NCCO) may have influenced the determination of the symmetry \cite{Cooper} from temperature-dependent penetration depth measurements \cite{Wu}.

In order to more definitively determine the pairing state symmetry of the electron-doped cuprates, we have again studied the low-temperature behavior of the penetration depth, $\lambda$(T).  The functional form of $\lambda$(T) is determined by the low-energy excitations of the system, and thus indirectly determines the pairing state symmetry of these materials.  In addition, we are now able to observe the behavior of the penetration depth to a lower temperature than was done previously \cite{Wu}.  This, along with the possible role of paramagnetism in previous measurements, makes it prudent to study an electron-doped system which is not strongly paramagnetic, such as the cuprate superconductor Pr$_{2-x}$Ce$_{x}$CuO$_{4-y}$ (PCCO).

The samples studied were crystals grown using a directional solidification technique and have been characterized in previous studies \cite{Wu}\cite{Peng}\cite{Anlage}.  Typically, these samples exhibited a transition temperature of 19 K for PCCO and 24 K for NCCO with a transition width in the surface resistance of 1.5 K and residual normal state resistivity values of about 60 $\mu\Omega$-cm.  A typical sample size was 1 mm x 1 mm x 30 $\mu$m.  The phase diagram for the electron-doped cuprates \cite{Maiser} shows that a cerium dopant concentration of x=0.15 has the highest T$_c$.  Based upon the resistivity, T$_c$, and transition width, we believe that our samples are at the optimal doping level.

\section{Experimental Method}

The in-plane penetration depth, $\lambda_{ab}$(T), as well as the surface resistance, R$_S$(T), were measured using a superconducting niobium microwave resonant cylindrical cavity operating at 9.6 GHz in which the TE$_{011}$ mode is stimulated.  In this mode, the magnetic field is maximum and the electric field zero at the center of the cavity.  The sample is placed on a hot finger in the center of the cavity with its c-axis aligned parallel to H$_{rf}$.  This induces screening currents in the copper-oxide planes of the crystal.  The sample temperature was varied from 1.2 K up to above T$_c$.  As the penetration depth changes with temperature, the resonant frequency, f(T), and quality factor, Q(T), of the cavity change.  By measuring these shifts in f(T) and Q(T), quantities such as the temperature dependence of $\Delta\lambda$(T) and R$_s$(T) can be deduced.  Further details of this technique are given elsewhere \cite{Anlage}.

There are some important improvements to this technique that are reported here for the first time.  In the past \cite{Wu}, four issues were of major concern regarding the microwave resonant cavity technique.  These involved the base temperature, the reproducibility of the background, the effects of changing liquid helium hydrostatic pressure on the dimensions of the resonator, and the orientation of the sample with respect to H$_{rf}$.

In our present design, we are able to reach a base temperature of 1.2 K without the use of an exchange gas, which can corrupt the data.  This has been accomplished by pumping on the helium bath and providing a stronger thermal link between the sample and the helium bath.  By reaching a lower base temperature, we are able to explore sample properties in a region of temperature in which the distinction between the different possible pairing state symmetries is most apparent.  Furthermore, it is at the lower temperatures where any paramagnetic influence would be most evident.

A further improvement was realized by permanently mounting the sample rod to the base of the apparatus.  To obtain useful data, it is necessary to introduce the sample to the proper position within the resonant cavity.  This position has a minimal E$_{rf}$, and the H$_{rf}$ is not only maximum, but aligned in a known direction, in our case, along the axis of the cylindrical cavity.  Furthermore, by having the sample rod permanently secured, we can ensure that the sample will be introduced to the same location within the cavity every time.  This is critical in order for the background measurements to be reproducible.

Also, we have addressed the issue of the helium hydrostatic pressure.  As the level of the liquid helium drops over the course of an experimental run, the pressure it exerts on the cavity decreases, causing the cavity to expand slightly.  This results in a constant drop in resonant frequency with time.  In order to minimize this effect, it is necessary to isolate the cavity from the helium bath, while still maintaining a good thermal link between the two.  To achieve this, we surround the resonant cavity with a vacuum jacket.  Thermal links were maintained by direct surface-to-surface contact between the cavity and this outer cylinder at certain points.  The result of this improvement was to reduce any drift in frequency due to changes in the helium hydrostatic pressure to less than the random noise present in the data.

Finally, the orientation of the sample in the cavity is different from previous experiments \cite{Wu}.  H$_{rf}$ is now applied parallel to the c-axis of the sample, whereas in the past, it was parallel to the a-b plane of the crystal.  This change has two effects.  First, the rare-earth paramagnetism is stronger when H$_{rf}$ is applied in the plane \cite{Dali}.  Thus, by changing the orientation we were able to significantly reduce the paramagnetic influence on our measurements.  Secondly, the former orientation induced c-axis currents in the sample.  This led to the simultaneous measurement of $\lambda_{ab}$(T) and $\lambda_c$(T), making it difficult to analyze the in-plane penetration depth without c-axis contributions.  However, with H$_{rf}$ now applied parallel to the c-axis, only a-b plane currents are induced.  This allows for a direct measurement of $\lambda_{ab}$(T).

\section{Results}

The surface resistance as a function of T/T$_c$ is presented in figure 1 for one PCCO and one NCCO crystal studied.  Note that both samples display a single transition at T$_c$.  Furthermore, the data have the same behavior until the PCCO crystal surface resistance saturates at the residual resistance value.  The homogeneity of the crystals is attested to by the nature of the transition.  Furthermore, the width of the transition, about 0.15 T$_c$, is typical for our samples.  This, along with their reasonable values of T$_c$, indicates that the doping is near optimal levels.  Therefore, the low temperature behavior of the penetration depth for these samples should be representative of the electron-doped cuprate superconductors, with minimal extrinsic influences.  This is in contrast to the behavior of other crystals which displayed much broader transitions as well as step-like features in R$_S$(T), indicative of additional superconducting transitions.  These crystals also showed multiple transitions in the c-axis resistivity, were judged to be inhomogeneous, and subsequently rejected.

The low temperature behavior of the change in penetration depth, $\Delta\lambda$(T), has been used as an indication of the pairing symmetry \cite{Wu}\cite{Anlage}.  In the case of BCS s-wave symmetry, $\Delta\lambda$(T) is expected to behave exponentially, the exact behavior dependent upon the zero-temperature penetration depth, $\lambda$(0), and the energy gap, $\Delta$(0).  The asymptotic BCS s-wave form is: $\Delta\lambda(T) = \lambda(0)(\pi\Delta(0)/2k_BT)^{1/2}e^{-\Delta(0)/k_BT}$ for T$<<$T$_c$/2.

In figure 2 we show $\Delta\lambda$(T) for an NCCO crystal and a PCCO crystal up to 0.45T$_c$, as well as the asymptotic BCS s-wave model using $\lambda$(0) = 1500 \AA\ and 2$\Delta(0)/k_BT_c$ = 3.50.  The data for the electron-doped crystals clearly differs from a simple BCS s-wave behavior, with the PCCO data exhibiting clear deviations from an exponential behavior at T/T$_c<$0.3.  Nevertheless, we can force the data to fit the asymptotic s-wave BCS form, given above, allowing both $\lambda(0)$ and $\Delta(0)$ to vary.  For the PCCO crystal we found $\lambda$(0) = 1800 \AA\ and 2$\Delta$(0)/k$_B$T$_c$ = 2.75.  Similarly, the NCCO crystal yields $\lambda$(0) = 2725 \AA\ and 2$\Delta$(0)/k$_B$T$_c$ = 2.35.  In both cases we find small gap values, much less than the 3.5 expected from BCS theory.  The gap values are also much smaller than previously reported in NCCO \cite{Wu}\cite{Anlage}\cite{Schn}\cite{Andr}.

Also of note in figure 2 is the upturn in the NCCO $\Delta\lambda$(T) data at low temperatures.  We believe that this upturn is due to the paramagnetism of the Nd$^{3+}$ ions, and its presence may explain why previous studies, which did not take this into account, found NCCO to be an s-wave superconductor.  In the fit considered above, we accounted for the paramagnetism by excluding data below 0.16 T$_c$ from our analysis.  At temperatures above 0.16 T$_c$, we believe the paramagnetic influence is less than in previous studies due to the difference in orientation of the sample.  It is important to realize that a similar paramagnetic upturn is not observed in the PCCO data due to the much weaker paramagnetism in this compound \cite{Dali}.  Thus, the PCCO data should more closely reflect the intrinsic behavior of the penetration depth.  A more detailed analysis of this data, and data on thin films, is forthcoming in which we conclude that our data fits best to a dirty d-wave model \cite{PRL}.

\begin{figure}
\begin{center}
\leavevmode
\epsfig{file=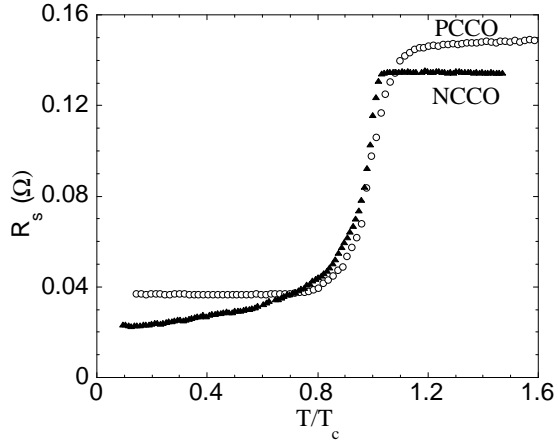,width=7.5cm,clip=}
\end{center}
\caption{Surface resistance, R$_S$(T/T$_c$), for PCCO (open circles) and NCCO (closed triangles) crystals, measured at 9.6 GHz, as a function of T/T$_c$.}
\label{schematic}
\end{figure}

\begin{figure}
\begin{center}
\leavevmode
\epsfig{file=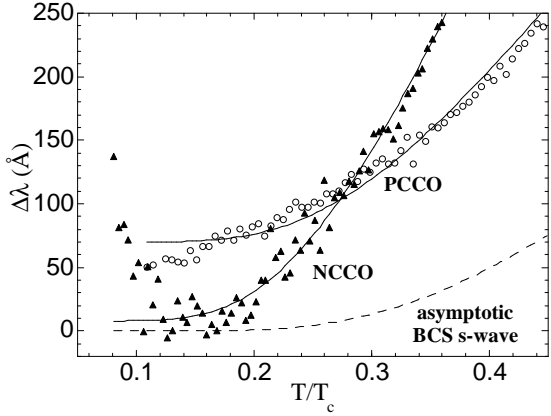,width=7.5cm,clip=}
\end{center}
\caption{$\Delta\lambda$(T/T$_c$) as a function of T/T$_c$ for crystals of PCCO (open circles), NCCO (closed triangles), and the asymptotic BCS s-wave model (dashed line).  The solid lines are fits to an asymptotic BCS s-wave of the form $\Delta\lambda(T) = \lambda(0)(\pi\Delta(0)/2k_BT)^{1/2}e^{-\Delta(0)/k_BT}$ Note: PCCO data is offset and NCCO data is fit for T$>$0.16T$_c$.}
\label{schematic}
\end{figure}

\section{Discussion}

Alff {\it et al.} recently obtained $\Delta\lambda$(T)/$\lambda$(0) data on NCCO and PCCO grain boundary junctions in thin films \cite{Alff}.  They concluded that their data was consistent with an s-wave scenario.  In the case of NCCO, they applied the same paramagnetic correction to their NCCO data which Cooper used to arrive at a d-wave explanation \cite{Cooper}.  This is evidence of how sensitive the final determination of the pairing state symmetry is to the choice of parameters for the Curie-Weiss law.  Although not shown here, we can also force our $\Delta\lambda$(T) data for a number of PCCO crystals to fit to an s-wave form.  In general however, this requires using unphysically large $\lambda$(0) and small $\Delta$(0) values.  Alff {\it et al.} reported a value for 2$\Delta(0)/k_BT_c$ of about 2.9.  This is less than the BCS value of 3.5, but compares well with our s-wave fit gap values.  However, we interpret this as a reason for rejecting an s-wave behavior.  Although both we and Alff {\it et al.} obtained very similar data, we believe that the paramater values exclude a standard BCS s-wave explanation.

However, there are some s-wave scenarios which might be viable.  For instance, gapless superconductivity could lead to T$^2$ behavior for the temperature dependent penetration depth at low temperatures.  Furthermore, the presence of magnetic impurities could potentially reduce the size of the activation gap.  This could then explain the small values for $\Delta$(0) we obtain when fitting to an s-wave functional form.  Nevertheless, the data is not consistent with a large-gap isotropic BCS s-wave behavior.

\section{Conclusion}

In conclusion, we have successfully employed a microwave resonant cavity system to measure the penetration depth and surface resistance of NCCO and PCCO crystals down to 1.2 K.  This data supports the conjecture that rare-earth paramagnetism might affect the observed penetration depth temperature dependence in NCCO.  By analyzing the functional form of this temperature dependence, we conclude that the pairing symmetry of the electron-doped cuprate superconductors is not a standard s-wave and is more likely d-wave in nature.

This work was supported by NSF DMR-9732736, NSF DMR-9624021, and the Maryland Center for Superconductivity Research.  We acknowledge discussions with D. H. Wu and R. Prozorov.

\end{document}